%% file: dlrm.tex
\documentclass{article}

\usepackage[nonatbib,preprint]{neurips_2019}

\usepackage[utf8]{inputenc} 
\usepackage[T1]{fontenc}    
\usepackage{hyperref}       
\usepackage{url}            
\usepackage{booktabs}       
\usepackage{amsfonts}       
\usepackage{nicefrac}       
\usepackage{microtype}      
\usepackage{graphicx}
\usepackage{subcaption}
\usepackage{amsmath}
\usepackage{algorithm}
\usepackage{algorithmic}
\usepackage{bm}
\usepackage{wrapfig}

\newcommand{\modfootnote}[1]{\footnote{\scriptsize{#1}}}
\newcommand{\dlrmlocation}{https://github.com/facebookresearch/dlrm}
\newcommand{\company}{large internet companies}
\newcommand{\squeeze}{\vspace{-10pt}}

\title{Deep Learning Recommendation Model for Personalization and Recommendation Systems}

\author{
  Maxim Naumov, Dheevatsa Mudigere, Hao-Jun Michael Shi\thanks{Northwestern University, $^\dagger$Harvard University, work done while at Facebook.}, Jianyu Huang, \\
  \textbf{Narayanan Sundaraman, Jongsoo Park, Xiaodong Wang, Udit Gupta$^\dagger$, Carole-Jean Wu,} \\
  \textbf{Alisson G. Azzolini, Dmytro Dzhulgakov, Andrey  Mallevich, Ilia Cherniavskii, Yinghai Lu,} \\
  \textbf{Raghuraman Krishnamoorthi, Ansha Yu, Volodymyr Kondratenko, Stephanie Pereira,} \\
  \textbf{Xianjie Chen, Wenlin Chen, Vijay Rao, Bill Jia, Liang Xiong and Misha Smelyanskiy} \\
  Facebook, 1 Hacker Way, Menlo Park, CA 94065 \\
  \texttt{\{mnaumov,dheevatsa\}@fb.com} \\
}

\begin{document}

\maketitle

\begin{abstract}

With the advent of deep learning, neural network-based recommendation models have emerged as an important tool for tackling personalization and recommendation tasks. These networks differ significantly from other deep learning networks due to their need to handle categorical features and are not well studied or understood. In this paper, we develop a state-of-the-art deep learning recommendation model (DLRM) and provide its implementation in both PyTorch and Caffe2 frameworks. In addition, we design a specialized parallelization scheme utilizing model parallelism on the embedding tables to mitigate memory constraints while exploiting data parallelism to scale-out compute from the fully-connected layers. We compare DLRM against existing recommendation models and characterize its performance on the Big Basin AI platform, demonstrating its usefulness as a benchmark for future algorithmic experimentation and system co-design. 

\end{abstract}

\input{introduction}

\input{model}
\input{parallelism}

\input{data}
\input{experiments}

\input{conclusion}

\subsubsection*{Acknowledgments}
The authors would like to acknowledge AI Systems Co-Design, Caffe2, PyTorch and AML team members for their help in reviewing this document.  

\small
\bibliographystyle{plain}
\bibliography{dlrm}

\end{document}

%% file: introduction.tex
\section{Introduction}

Personalization and recommendation systems are currently deployed for a variety of tasks at \company, including ad click-through rate (CTR) prediction and rankings. Although these methods have had long histories, these approaches have only recently embraced neural networks. Two primary perspectives contributed towards the architectural design of deep learning models for personalization and recommendation.

The first comes from the view of recommendation systems. These systems initially employed content filtering where a set of experts classified products into categories, while users selected their preferred categories and were matched based on their preferences \cite{mgp}. The field subsequently evolved to use collaborative filtering, where recommendations are based on past user behaviors, such as prior ratings given to products. Neighborhood methods \cite{ning2015} that provide recommendations by grouping users and products together and latent factor methods that characterize users and products by certain implicit factors via matrix factorization techniques \cite{frolov2017tensor, koren2009matrix} were later deployed with success. 

The second view comes from predictive analytics, which relies on statistical models to classify or predict the probability of events based on the given data \cite{Devroye1996}. Predictive models shifted from using simple models such as linear and logistic regression \cite{walker1967} to models that incorporate deep networks. In order to process categorical data, these models adopted the use of embeddings, which transform the one- and multi-hot vectors into dense representations in an abstract space \cite{naumov2019}. This abstract space may be interpreted as the space of the latent factors found by recommendation systems.

In this paper, we introduce a personalization model that was conceived by the union of the two perspectives described above. The model uses embeddings to process sparse features that represent categorical data and a multilayer perceptron (MLP) to process dense features, then interacts these features explicitly using the statistical techniques proposed in \cite{rendle2010}. Finally, it finds the event probability by post-processing the interactions with another MLP. We refer to this model as a deep learning recommendation model (DLRM); see Fig. \ref{fig:ranking_model}. A PyTorch and Caffe2 implementation of this model will be released for testing and experimentation with the publication of this manuscript.

%% file: model.tex
\section{Model Design and Architecture}

In this section, we will describe the design of DLRM. We will begin with the high level components of the network and explain how and why they have been assembled together in a particular way, with implications for future model design, then characterize the low level operators and primitives that make up the model, with implications for future hardware and system design.

\subsection{Components of DLRM}

\begin{wrapfigure}{r}{0.5\textwidth}
    \centering
    \includegraphics[width=0.4\columnwidth]{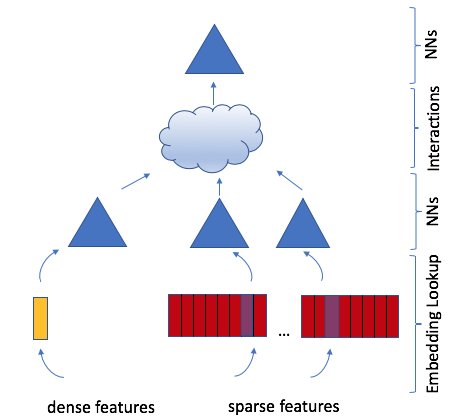}
    \caption{A deep learning recommendation model}
    \label{fig:ranking_model}
\end{wrapfigure}

The high-level components of the DLRM can be more easily understood by reviewing early models. We will avoid the full scientific literature review and focus instead on the four techniques used in early models that can be interpreted as salient high-level components of the DLRM.

\subsubsection{Embeddings}

In order to handle categorical data, embeddings map each category to a dense representation in an abstract space. In particular, each embedding lookup may be interpreted as using a one-hot vector $\bm{e}_i$ (with the $i$-th position being $1$ while others are $0$, where index $i$ corresponds to $i$-th category) to obtain the corresponding row vector of the embedding table $W \in \mathbb{R}^{m \times d}$ as follows
\begin{equation}
\bm{w}_i^{T} = \bm{e}_i^{T} W .  
\end{equation} 
In more complex scenarios, an embedding can also represent a weighted combination of multiple items, with a multi-hot vector of weights $\bm{a}^T = [0,...,a_{i_1},...,a_{i_k},...,0]$, with elements $a_{i} \ne 0$ for $i = i_1, ..., i_k$ and $0$ everywhere else, where $i_1, ..., i_k$ index the corresponding items. Note that a mini-batch of $t$ embedding lookups can hence be written as
\begin{equation}
    S = A^{T} W
    \label{eq:batchedlookup} 
\end{equation}
where sparse matrix $A = [\bm{a}_1,...,\bm{a}_t]$ \cite{naumov2019}. 

DLRMs will utilize embedding tables for mapping categorical features to dense representations. However, even after these embeddings are meaningfully devised, how are they to be exploited to produce accurate predictions? To answer this, we return to latent factor methods. 

\subsubsection{Matrix Factorization}

Recall that in the typical formulation of the recommendation problem, we are given a set $\mathcal{S}$ of users that have rated some products. We would like to represent the $i$-th product by a vector $\bm{w}_i \in \mathbb{R}^d$ for $i = 1, ..., n$ and $j$-th user by a vector $\bm{v}_j \in \mathbb{R}^d$ for $j = 1, ..., m$ to find all the ratings, where $n$ and $m$ denote the total number of products and users, respectively. More rigorously, the set $\mathcal{S}$ consists of tuples $(i, j)$ indexing when the $i$-th product has been rated by the $j$-th user.

The matrix factorization approach solves this problem by minimizing
\begin{equation}
    \min \sum_{(i,j) \in \mathcal{S}} r_{ij} - \bm{w}_i^{T}\bm{v}_{j}
    \label{eq:matrixfactorization}
\end{equation}
where $r_{ij} \in \mathbb{R}$ is the rating of the $i$-th product by the $j$-th user for $i = 1, ... ,m$ and $j = 1, ..., n$. Then, letting $W^T=[\bm{w}_1,...,\bm{w}_m]$ and $V^T=[\bm{v}_1,...,\bm{v}_n]$, we may approximate the full matrix of ratings $R=[r_{ij}]$ as the matrix product $R \approx WV^T$. Note that $W$ and $V$ may be interpreted as two embedding tables, where each row represents a user/product in a latent factor space\modfootnote{This problem is different from low-rank approximation, which can be solved by SVD \cite{golub1996}, because not all entries of matrix $R$ are known.} \cite{koren2009matrix}. The dot product of these embedding vectors yields a meaningful prediction of the subsequent rating, a key observation to the design of factorization machines and DLRM.
 
\subsubsection{Factorization Machine}

In classification problems, we want to define a prediction function $\phi: \mathbb{R}^{n} \rightarrow T$ from an input datapoint $\bm{x} \in \mathbb{R}^{n}$ to a target label $y \in T$. As an example, we can predict the click-through rate by defining $T=\{+1,-1\}$ with $+1$ denoting the presence of a click and $-1$ as the absence of a click.

Factorization machines (FM) incorporate second-order interactions into a linear model with categorical data by defining a model of the form
\begin{equation}
    \hat{y} = b + \bm{w}^T\bm{x} + \bm{x}^T \texttt{upper}(VV^T)\bm{x}
    \label{eq:fm}
\end{equation}
where $V \in \mathbb{R}^{n \times d}$, $\bm{w} \in \mathbb{R}^n$, and $b \in \mathbb{R}$ are the parameters with $d \ll n$, and \texttt{upper} selects the strictly upper triangular part of the matrix \cite{rendle2010}. 

FMs are notably distinct from support vector machines (SVMs) with polynomial kernels \cite{Corinna1995} because they factorize the second-order interaction matrix into its latent factors (or embedding vectors) as in matrix factorization, which more effectively handles sparse data. This significantly reduces the complexity of the second-order interactions by only capturing interactions between pairs of distinct embedding vectors, yielding linear computational complexity.

\subsubsection{Multilayer Perceptrons}
Simultaneously, much recent success in machine learning has been due to the rise of deep learning. The most fundamental model of these is the multilayer perceptron (MLP), a prediction function composed of an interleaving sequence of fully connected (FC) layers and an activation function $\sigma : \mathbb{R} \rightarrow \mathbb{R}$ applied componentwise as shown below
\begin{equation}
    \hat{y} = W_k \sigma( W_{k - 1} \sigma( ... \sigma( W_1 \bm{x} + \bm{b}_1 ) ...) + \bm{b}_{k - 1} ) + \bm{b}_k
    \label{eq:mlp}
\end{equation}
where weight matrix $W_l \in \mathbb{R}^{n_l \times n_{l - 1}}$, bias $\bm{b}_l \in \mathbb{R}^{n_l}$ for layer $l=1,...,k$. 

These methods have been used to capture more complex interactions. It has been shown, for example, that given enough parameters, MLPs with sufficient depth and width can fit data to arbitrary precision \cite{bishop1995}. Variations of these methods have been widely used in various applications including computer vision and natural language processing. One specific case, Neural Collaborative Filtering (NCF) \cite{he2017neural,sedhain2015autorec} used as part of the MLPerf benchmark \cite{mlperf}, uses an MLP rather than dot product to compute interactions between embeddings in matrix factorization.

\subsection{DLRM Architecture}

So far, we have described different models used in recommendation systems and predictive analytics. Let us now combine their intuitions to build a state-of-the-art personalization model. 

Let the users and products be described by many continuous and categorical features. To process the categorical features, each categorical feature will be represented by an embedding vector of the same dimension, generalizing the concept of latent factors used in matrix factorization \eqref{eq:matrixfactorization}. To handle the continuous features, the continuous features will be transformed by an MLP (which we call the \textit{bottom} or \textit{dense} MLP) which will yield a dense representation of the same length as the embedding vectors \eqref{eq:mlp}.

We will compute second-order interaction of different features explicitly, following the intuition for handling sparse data provided in FMs \eqref{eq:fm}, optionally passing them through MLPs. This is done by taking the dot product between all pairs of embedding vectors and processed dense features. These dot products are concatenated with the original processed dense features and post-processed with another MLP (the \textit{top} or \textit{output} MLP) \eqref{eq:mlp}, and fed into a sigmoid function to give a probability.

We refer to the resulting model as DLRM, shown in Fig. \ref{fig:ranking_model}. We show some of the operators used in DLRM in PyTorch \cite{paszke2017pytorch} and Caffe2 \cite{caffe2}  frameworks in Table \ref{tab:ops}.

\squeeze

\begin{table}[h]
    \centering \small
    \begin{tabular}{c|c|c|c|c}
            & Embedding                & MLP                      & Interactions    & Loss         \\ \hline
    PyTorch & \texttt{nn.EmbeddingBag} & \texttt{nn.Linear/addmm} & \texttt{matmul/bmm} & \texttt{nn.CrossEntropyLoss}\\ \hline
    Caffe2  & \texttt{SparseLengthSum} & \texttt{FC}         & \texttt{BatchMatMul} & \texttt{CrossEntropy} \\ 
    \end{tabular}
    \caption{DLRM operators by framework}
    \label{tab:ops}
    \squeeze
\end{table}

\squeeze

\subsection{Comparison with Prior Models}

Many deep learning-based recommendation models \cite{cheng2016wide,guo2017deepfm,wang2017deep,lian2018xdeepfm,zhou2018deepi,zhou2018deep} use similar underlying ideas to generate higher-order terms to handle sparse features. Wide and Deep, Deep and Cross, DeepFM, and xDeepFM networks, for example, design specialized networks to systematically construct higher-order interactions. These networks then sum the results from both their specialized model and an MLP, passing this through a linear layer and sigmoid activation to yield a final probability. DLRM specifically interacts embeddings in a structured way that mimics factorization machines to significantly reduce the dimensionality of the model by only considering cross-terms produced by the dot-product between pairs of embeddings in the final MLP. We argue that higher-order interactions beyond second-order found in other networks may not necessarily be worth the additional computational/memory cost. 

A key difference between DLRM and other networks is in how these networks treat embedded feature vectors and their cross-terms. In particular, DLRM (and xDeepFM \cite{lian2018xdeepfm}) interpret each feature vector as a single unit representing a single category, whereas networks like Deep and Cross treat each element in the feature vector as a new unit that should yield different cross-terms. Hence, Deep and Cross networks will produce cross-terms not only between elements from different feature vectors as in DLRM via the dot product, but also produce cross-terms between elements within the same feature vector, resulting in higher dimensionality.

%% file: parallelism.tex
\section{Parallelism}
\label{sec:parallel}

Modern personalization and recommendation systems require large and complex models to capitalize on vast amounts of data. DLRMs particularly contain a very large number of parameters, up to multiple orders of magnitude more than other common deep learning models like convolutional neural networks (CNN), transformer and recurrent networks (RNN), and generative networks (GAN). This results in training times up to several weeks or more. Hence, it is important to parallelize these models efficiently in order to solve these problems at practical scales. 

As described in the previous section, DLRMs process both categorical features (with embeddings) and continuous features (with the bottom MLP) in a coupled manner. Embeddings contribute the majority of the parameters, with several tables each requiring in excess of multiple GBs of memory, making DLRM memory-capacity and bandwidth intensive. The size of the embeddings makes it prohibitive to use data parallelism since it requires replicating large embeddings on every device. In many cases, this memory constraint necessitates the distribution of the model across multiple devices to be able satisfy memory capacity requirements. 

On the other hand, the MLP parameters are smaller in memory but translate into sizeable amounts of compute. Hence, data-parallelism is preferred for MLPs since this enables concurrent processing of the samples on different devices and only requires communication when accumulating updates. Our parallelized DLRM will use a combination of model parallelism for the embeddings and data parallelism for the MLPs to mitigate the memory bottleneck produced by the embeddings while parallelizing the forward and backward propagations over the MLPs. Combined model and data parallelism is a unique requirement of DLRM as a result of its architecture and large model sizes. Such combined parallelism is not supported in either Caffe2 or PyTorch (as well as other popular deep learning frameworks), therefore we design a custom implementation. We plan to provide its detailed performance study in forthcoming work.

\begin{figure}[h!]
    \centering
    \includegraphics[width=0.65\columnwidth]{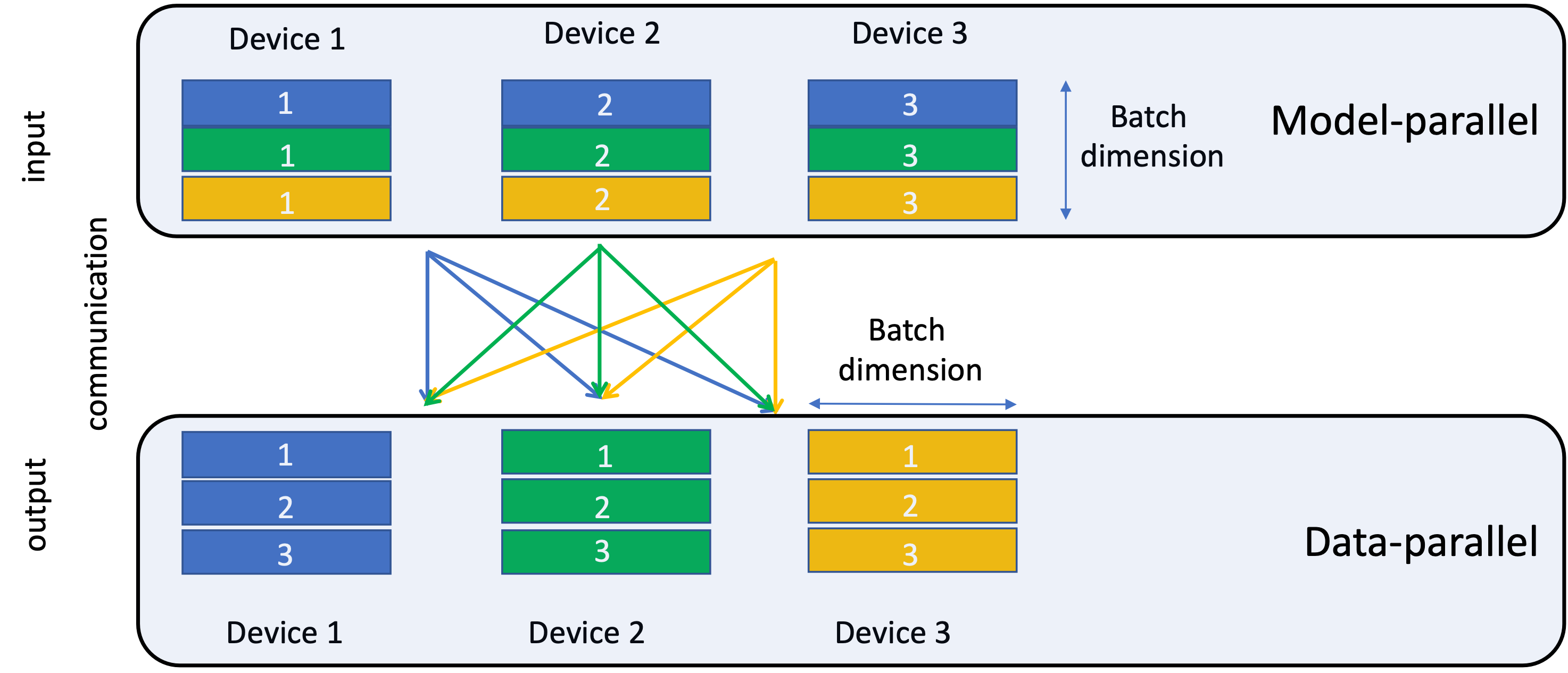}
    \caption{Butterfly shuffle for the all-to-all (personalized) communication}
    \label{fig:a2aP}
    \squeeze
\end{figure}

In our setup, the top MLP and the interaction operator require access to part of the mini-batch from the bottom MLP and all of the embeddings. Since model parallelism has been used to distribute the embeddings across devices, this requires a personalized all-to-all communication \cite{commsref}. At the end of the embedding lookup, each device has a vector for the embedding tables resident on those devices for all the samples in the mini-batch, which needs to be split along the mini-batch dimension and communicated to the appropriate devices, as shown in Fig. \ref{fig:a2aP}. Neither PyTorch nor Caffe2 provide native support for model parallelism; therefore, we have implemented it by explicitly mapping the embedding operators (\texttt{nn.EmbeddingBag} for PyTorch, \texttt{SparseLengthSum} for Caffe2) to different devices. Then personalized all-to-all communication is implemented using the butterfly shuffle operator, which appropriately slices the resulting embedding vectors and transfers them to the target devices. In the current version, these transfers are explicit copies, but we intend to further optimize this using the available communication primitives (such as \texttt{all-gather} and \texttt{send-recv}).

We note that for the data parallel MLPs, the parameter updates in the backward pass are accumulated with an \texttt{allreduce}\modfootnote{Optimized implementations for the \texttt{allreduce} op. include Nvidia's NCCL \cite{nccl} and Facebook's gloo \cite{gloo}.} and applied to the replicated parameters on each device \cite{commsref} in a synchronous fashion, ensuring the updated parameters on each device are consistent before every iteration.  In PyTorch, data parallelism is enabled through the \texttt{nn.DistributedDataParallel} and \texttt{nn.DataParallel} modules that replicate the model on each device and insert \texttt{allreduce} with the necessary dependencies. In Caffe2, we manually insert \texttt{allreduce} before the gradient update.

%% file: data.tex
\section{Data}
In order to measure the accuracy of the model, test its overall performance, and characterize the individual operators, we need to create or obtain a data set for our implementation. Our current implementation of the model supplies three types of data sets: random, synthetic and public data sets. 

The former two data sets are useful in experimenting with the model from the systems perspective. In particular, it permits us to exercise different hardware properties and bottlenecks by generating data on the fly while removing dependencies on data storage systems. The latter allows us to perform experiments on real data and measure the accuracy of the model. 


\subsection{Random}
Recall that DLRM accepts continuous and categorical features as inputs. The former can be modeled by generating a vector of random numbers using either a uniform or normal (Gaussian) distributions with the \texttt{numpy.random} package \texttt{rand} or \texttt{randn} calls with default parameters. Then a mini-batch of inputs can be obtained by generating a matrix where each row corresponds to an element in the mini-batch.

To generate categorical features, we need to determine how many non-zero elements we would like have in a given multi-hot vector. The benchmark allows this number to be either fixed or random within a range\modfootnote{see options \texttt{-}\texttt{-num-indices-per-lookup=k} and \texttt{-}\texttt{-num-indices-per-lookup-fixed}} $[1,k]$. Then, we generate the corresponding number of integer indices, within a range $[1,m]$, where $m$ is the number of rows in the embedding $W$ in \eqref{eq:batchedlookup}.  Finally, in order to create a mini-batch of lookups, we concatenate the above indices and delineate each individual lookup with lengths (\texttt{SparseLengthsSum}) or offsets (\texttt{nn.EmbeddingBag})\modfootnote{For instance, in order to represent three embedding lookups, with indices $\{0,2\}$, $\{0,1,5\}$ and $\{3\}$ we use 
\begin{eqnarray}
    \texttt{lengths/offsets} &=& \{2,3,1\} / \{0,2,5\} \nonumber \\
    \texttt{indices} &=& \{0,2,0,1,5,3\} \nonumber
\end{eqnarray}
Note that this format resembles Compressed-Sparse Row (CSR) often used for sparse matrices in linear algebra.}.

\subsection{Synthetic}

There are many reasons to support custom generation of indices corresponding to categorical features. For instance, if our application uses a particular data set, but we would not like to share it for privacy purposes, then we may choose to express the categorical features through distributions. This could potentially serve as an alternative to the privacy preserving techniques used in applications such as federated learning \cite{bonawitz2019federated,gentry2009}. Also, if we would like to exercise system components, such as studying memory behavior, we may want to capture fundamental locality of accesses of original trace within synthetic trace.    

Let us now illustrate how we can use a synthetic data set. Assume that we have a trace of indices that correspond to embedding lookups for a single categorical feature (and repeat the process for all features). We can record the unique accesses and frequency of distances between repeated accesses in this trace (Alg. \ref{alg:profile}) and then generate a synthetic trace (Alg. \ref{alg:generate}) as proposed in \cite{hassan2007synthetic}. 

\begin{algorithm}[h]
    \caption{Profile (Original) Trace}
    \label{alg:profile}
    \begin{algorithmic}[1]
        \STATE Let \texttt{tr} be input sequence, \texttt{s} stack of distances, \texttt{u} list of unique accesses and \texttt{p} probability distribution 
        \STATE Let \texttt{s.position\_from\_the\_top} return $d=0$ if the index is not found, and $d>0$ otherwise.
        \FOR{\texttt{i=0}; \texttt{i<length(tr)}; \texttt{i++}}
            \STATE \texttt{a = tr[i]}  
            \STATE \texttt{d = s.position\_from\_the\_top(a)}
            \IF{ \texttt{d == 0}} 
                \STATE \texttt{u.append(a)}
            \ELSE                 
                \STATE \texttt{s.remove\_from\_the\_top\_at\_position(d)}
            \ENDIF                         
            \STATE \texttt{p[d]} += \texttt{1.0/length(tr)}  
            \STATE \texttt{s.push\_to\_the\_top(a)} 
        \ENDFOR
    \end{algorithmic}
\end{algorithm}
\squeeze
\begin{algorithm}[h]
    \caption{Generate (Synthetic) Trace}
    \label{alg:generate}
    \begin{algorithmic}[1]
        \STATE Let \texttt{u} be input list of unique accesses and \texttt{p} probability distribution of distances, while \texttt{tr} output trace. 
        \FOR{\texttt{s=0}, \texttt{i=0}; \texttt{i<length}; \texttt{i++}}
            \STATE \texttt{d = p.sample\_from\_distribution\_with\_support(0,s)}
            \IF{\texttt{d == 0}} 
                \STATE \texttt{a = u.remove\_from\_front()}
                \STATE \texttt{s++}
            \ELSE
                \STATE \texttt{a = u.remove\_from\_the\_back\_at\_position(d)} 
            \ENDIF
            \STATE \texttt{u.append(a)}
            \STATE \texttt{tr[i] = a}
        \ENDFOR
    \end{algorithmic}
\end{algorithm}

\begin{figure}[h!]
    \centering
    \begin{subfigure}[b]{.33\linewidth}
        \includegraphics[width=\linewidth,height=0.1245\textheight]{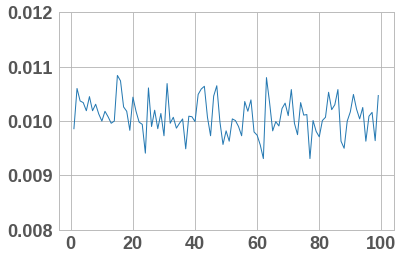}
        \caption{original}
    \end{subfigure}
    \hspace{-10pt}
    \begin{subfigure}[b]{.33\linewidth}
        \includegraphics[width=\linewidth,height=0.1245\textheight]{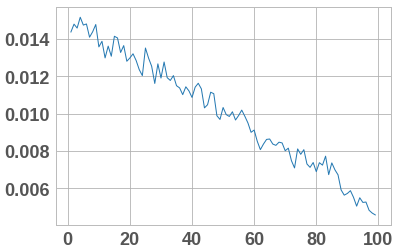}
        \caption{synthetic trace}
    \end{subfigure}
    \hspace{-10pt}
    \begin{subfigure}[b]{.33\linewidth}
        \includegraphics[width=\linewidth,height=0.1245\textheight]{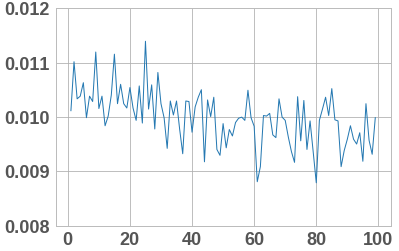}
        \caption{adjusted synthetic trace}
    \end{subfigure}
    \caption{Probability distribution \texttt{p} based on a sample trace  \texttt{tr = random.uniform(1,100,100K)}}
    \label{fig:synthetic_distributions}
    \squeeze
\end{figure}

Note that we can only generate a stack distance up to \texttt{s} number of unique accesses we have seen so far, therefore \texttt{s} is used to control the support of the distribution \texttt{p} in Alg. \ref{alg:generate}. Given a fixed number of unique accesses, the longer input trace will result in lower probability being assigned to them in Alg. \ref{alg:profile}, which will lead to longer time to achieve full distribution support in Alg. \ref{alg:generate}. In order to address this problem, we increase the probability for the unique accesses up to a minimum threshold and adjust support to remove unique accesses from it once all have been seen.  A visual comparison of probability distribution \texttt{p} based on original and synthetic traces is shown in Fig. \ref{fig:synthetic_distributions}. In our experiments original and adjusted synthetic traces produce similar cache hit/miss rates.

Alg. \ref{alg:profile} and \ref{alg:generate} were designed for more accurate cache simulations, but they illustrate a general idea of how probability distributions can be used to generate synthetic traces with desired properties.

\subsection{Public}

Few public data sets are available for recommendation and personalization systems. The Criteo AI Labs Ad Kaggle\modfootnote{https://www.kaggle.com/c/criteo-display-ad-challenge} and Terabyte\modfootnote{https://labs.criteo.com/2013/12/download-terabyte-click-logs/} data sets are open-sourced data sets consisting of click logs for ad CTR prediction. Each data set contains 13 continuous and 26 categorical features. Typically the continuous features are pre-processed with a simple log transform $\log(1 + x)$. The categorical feature are mapped to its corresponding embedding index, with unlabeled categorical features or labels mapped to 0 or \texttt{NULL}. 

The Criteo Ad Kaggle data set contains approximately 45 million samples over 7 days. In experiments, typically the 7th day is split into a validation and test set while the first 6 days are used as the training set. The Criteo Ad Terabyte data set is sampled over 24 days, where the 24th day is split into a validation and test set and the first 23 days is used as a training set. Note that there are an approximately equal number of samples from each day. 

%% file: experiments.tex
\section{Experiments}
\label{sec:exp}
\begin{wrapfigure}{r}{0.35\textwidth}
    \centering
    \includegraphics[width=0.275\columnwidth]{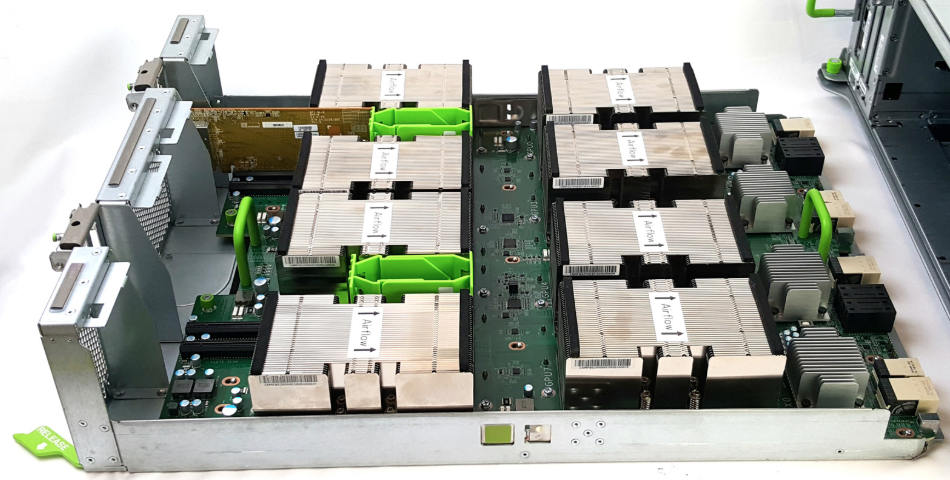}
    \caption{Big Basin AI platform}
    \label{fig:bigbasin}
    \squeeze    
\end{wrapfigure}
Let us now illustrate the performance and accuracy of DLRM. The model is implemented in PyTorch and Caffe2 frameworks and is available on \texttt{GitHub}\modfootnote{\dlrmlocation}. It uses \texttt{fp32} floating point and \texttt{int32}(Caffe2)/\texttt{int64}(PyTorch) types for model parameters and indices, respectively. The experiments are performed on the Big Basin platform with Dual Socket Intel Xeon 6138 CPU @ 2.00GHz and eight Nvidia Tesla V100 16GB GPUs, publicly available through the Open Compute Project\modfootnote{https://www.opencompute.org}, shown in Fig. \ref{fig:bigbasin}.

\subsection{Model Accuracy on Public Data Sets}

We evaluate the accuracy of the model on Criteo Ad Kaggle data set and compare the performance of DLRM against a Deep and Cross network (DCN) as-is without extensive tuning \cite{wang2017deep}. We compare with DCN because it is one of the few models that has comprehensive results on the same data set. Notice that in this case the models are sized to accommodate the number of features present in the data set. In particular, DLRM consists of both a bottom MLP for processing dense features consisting of three hidden layers with $512$, $256$ and $64$ nodes, respectively, and a top MLP consisting of two hidden layers with $512$ and $256$ nodes. On the other hand DCN consists of six cross layers and a deep network with $512$ and $256$ nodes. An embedding dimension of $16$ is used. Note that this yields a DLRM and DCN both with approximately $540M$ parameters.

\begin{figure}[ht]
    \centering
    \begin{subfigure}[b]{.4\linewidth}
        \includegraphics[width=\linewidth]{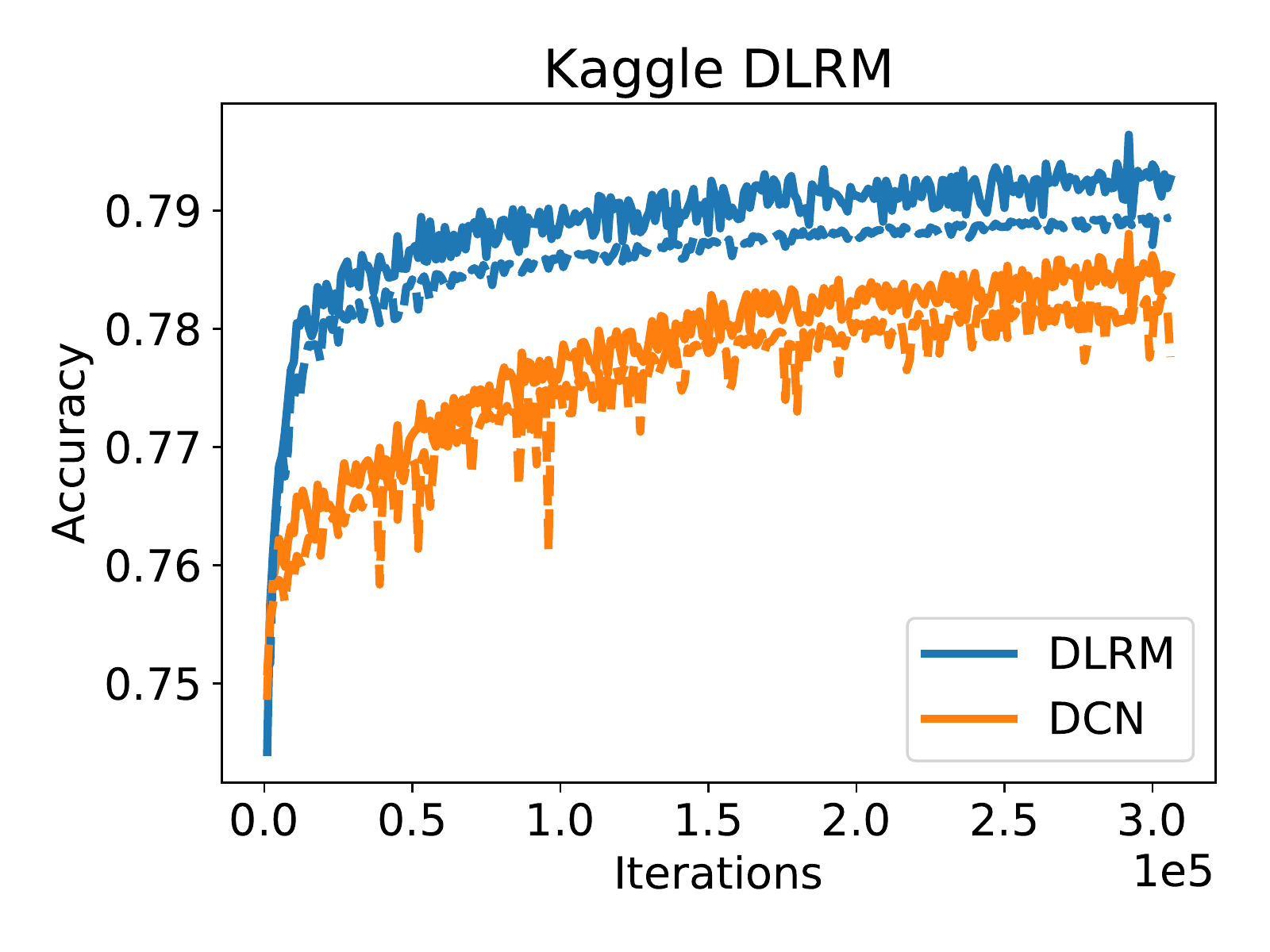}
        \squeeze
        \subcaption{SGD}
    \end{subfigure}
    \hspace{+10pt}
    \begin{subfigure}[b]{.4\linewidth}
        \includegraphics[width=\linewidth]{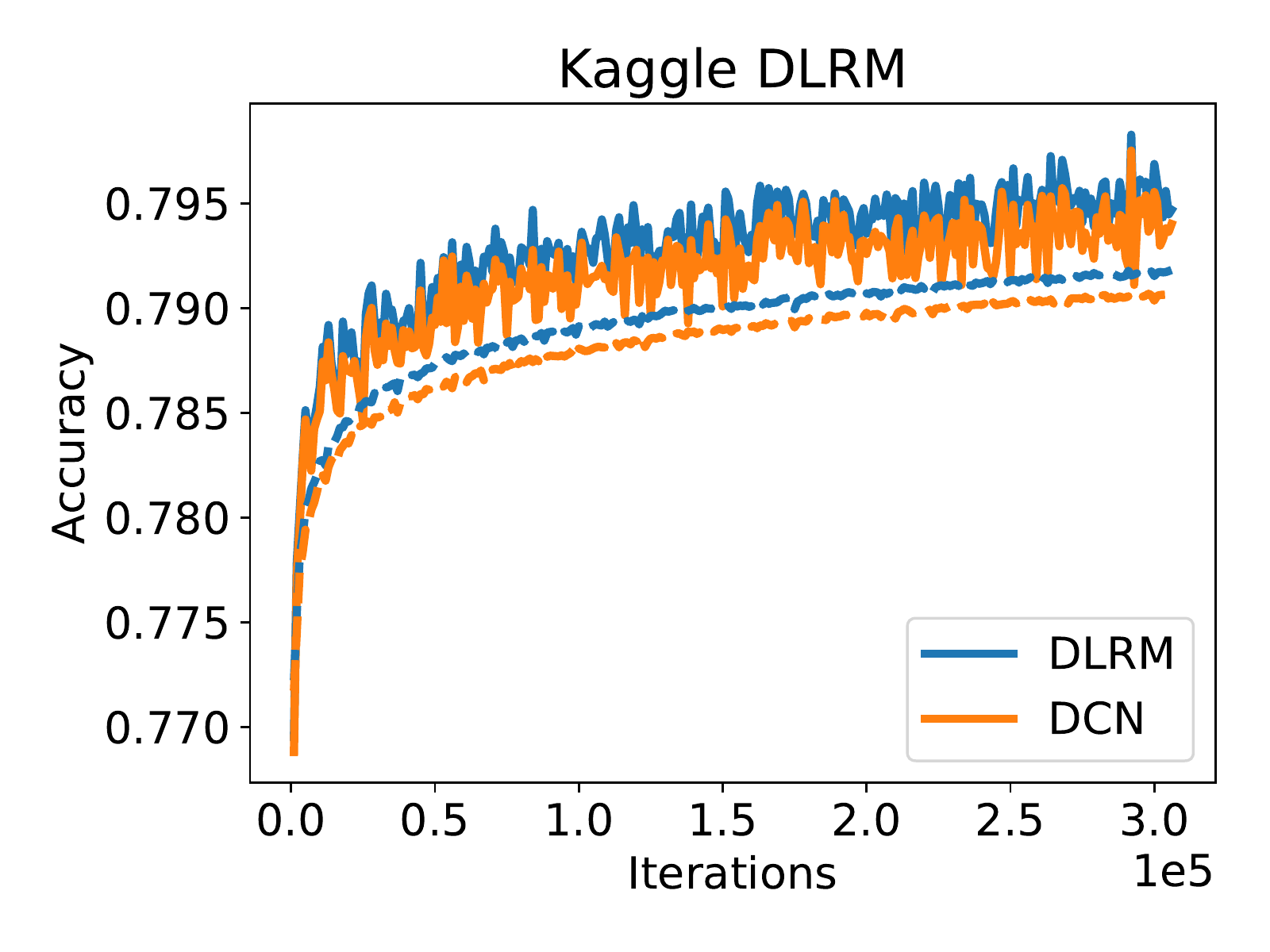}
        \squeeze
        \subcaption{Adagrad}
    \end{subfigure}
    \caption{Comparison of training (solid) and validation (dashed) accuracies of DLRM and DCN}
    \label{fig:kaggle dcn}
    \squeeze
\end{figure}

We plot both the training (solid) and validation (dashed) accuracies over a full single epoch of training for both models with SGD and Adagrad optimizers \cite{duchi2011}. No regularization is used. In this experiment, DLRM obtains slightly higher training and validation accuracy, as shown in Fig. \ref{fig:kaggle dcn}. We emphasize that this is without extensive tuning of model hyperparameters.

\subsection{Model Performance on a Single Socket/Device}

To profile the performance of our model on a single socket device, we consider a sample model with $8$ categorical features and $512$ continuous features. Each categorical feature is processed through an embedding table with $1M$ vectors, with vector dimension $64$, while the continuous features are assembled into a vector of dimension $512$. Let the bottom MLP have two layers, while the top MLP has four layers. We profile this model on a data set with $2048K$ randomly generated samples organized into $1K$ mini-batches\modfootnote{
For instance, this configuration can be achieved with the following command line arguments \\
\texttt{\texttt{-}-arch-embedding-size=1000000-1000000-1000000-1000000-1000000-1000000-1000000-1000000 \texttt{-}-arch-sparse-feature-size=64 \texttt{-}-arch-mlp-bot=512-512-64 \texttt{-}-arch-mlp-top=1024-1024-1024-1 \texttt{-}-data-generation=random \texttt{-}-mini-batch-size=2048 \texttt{-}-num-batches=1000 \texttt{-}-num-indices-per-lookup=100  [\texttt{-}-use-gpu] [\texttt{-}-enable-profiling]}}. 

\begin{figure}[h]
    \centering
    \begin{subfigure}[b]{.45\linewidth}
        \includegraphics[width=\linewidth,height=0.2\textheight]{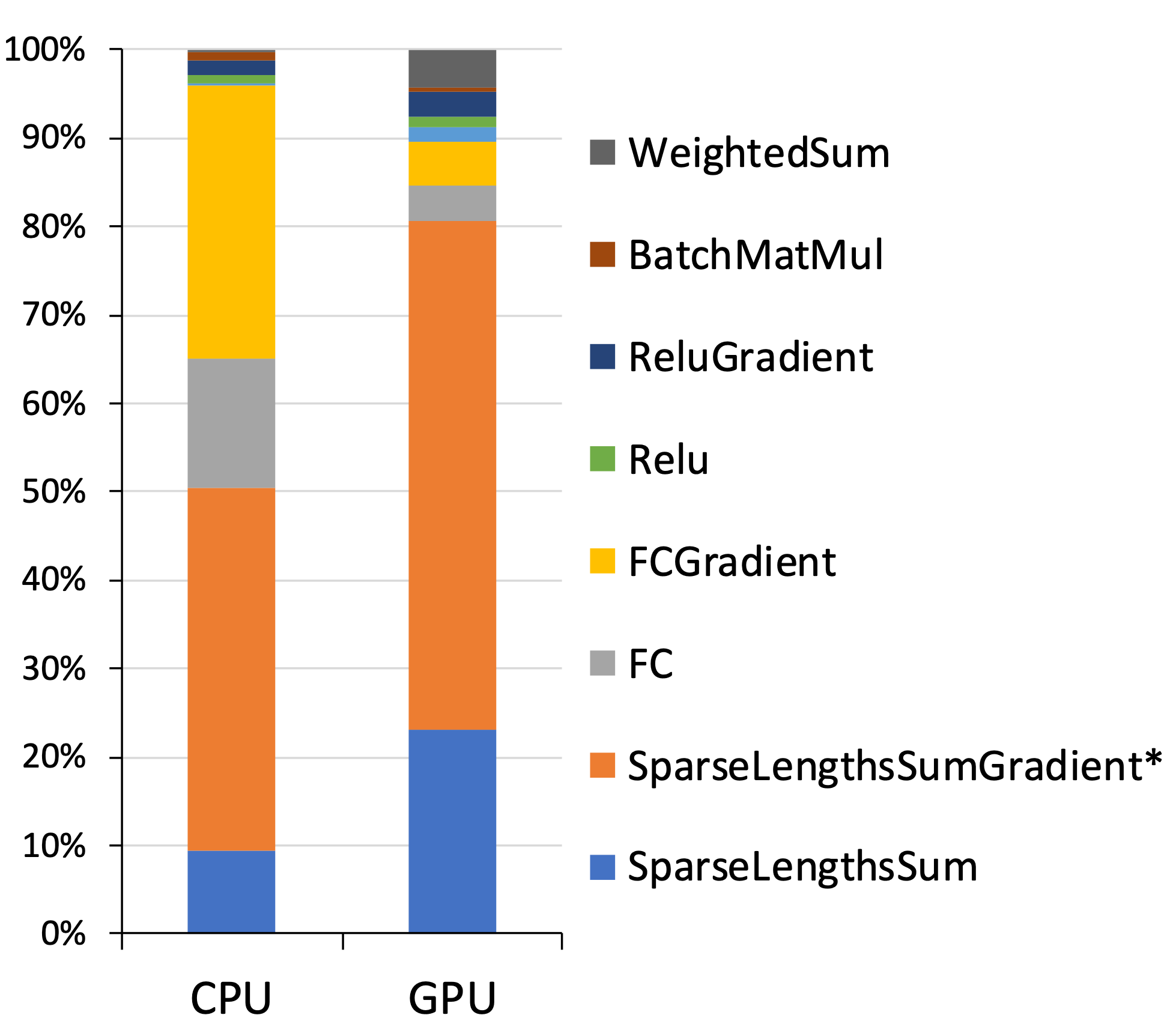}
        \squeeze
        \subcaption{Caffe2}
    \end{subfigure}
    \hspace{+10pt}
    \begin{subfigure}[b]{.45\linewidth}
        \includegraphics[width=\linewidth,,height=0.2\textheight]{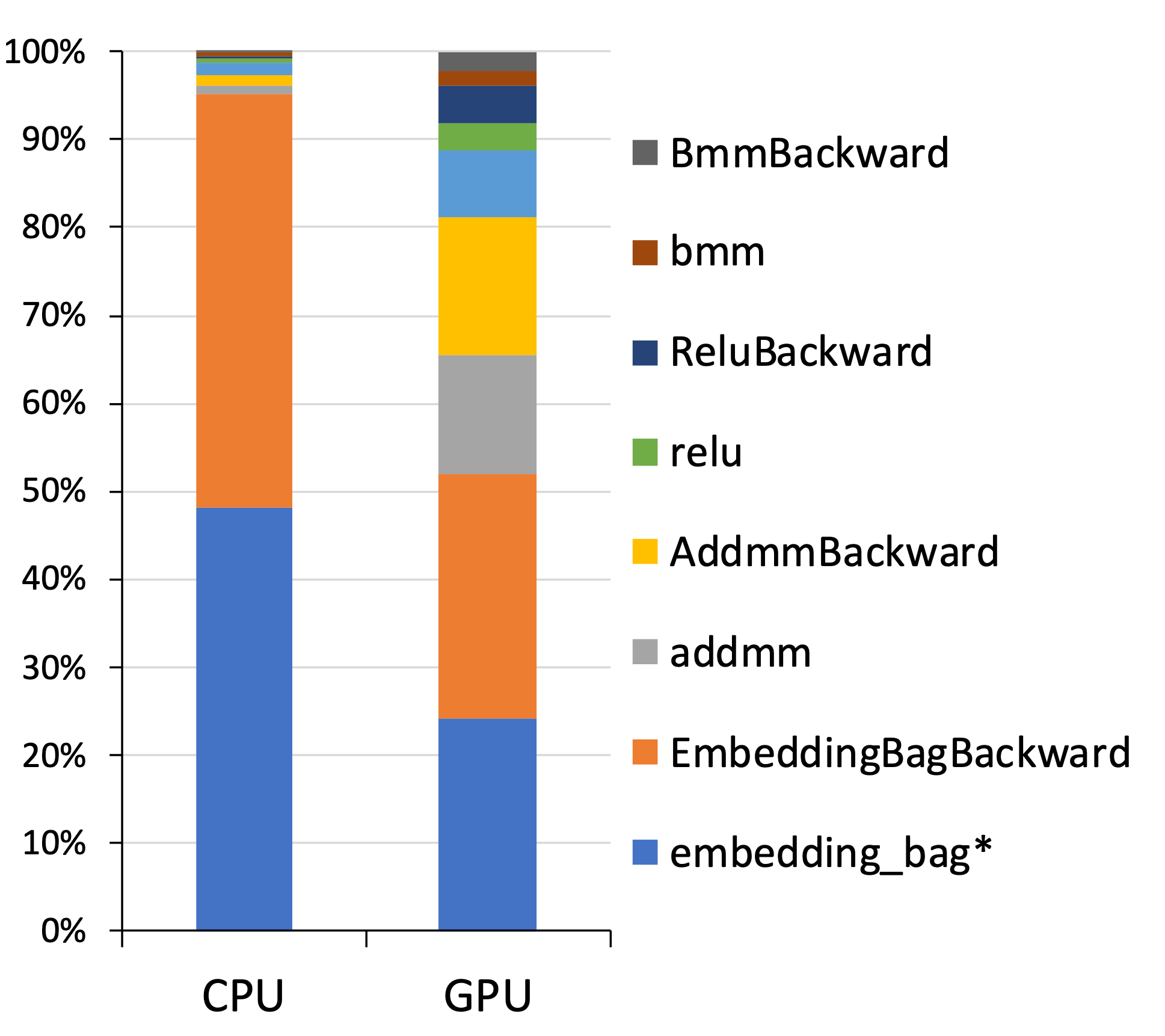}
        \squeeze
        \subcaption{PyTorch}
    \end{subfigure}
    \caption{Profiling of a sample DLRM on a single socket/device}
    \label{fig:profilingsingle}
    \squeeze
\end{figure}

This model implementation in Caffe2 runs in around 256 seconds on the CPU and 62 seconds on the GPU, with profiling of individual operators shown in Fig. \ref{fig:profilingsingle}. As expected, the majority of time is spent performing embedding lookups and fully connected layers. On the CPU, fully connected layers take a significant portion of the computation, while on the GPU they are almost negligible.

%% file: conclusion.tex
\section{Conclusion}

In this paper, we have proposed and open-sourced a novel deep learning-based recommendation model that exploits categorical data. Although recommendation and personalization systems still drive much practical success of deep learning within industry today, these networks continue to receive little attention in the academic community. By providing a detailed description of a state-of-the-art recommendation system and its open-source implementation, we hope to draw attention to the unique challenges that this class of networks present in an accessible way for the purpose of further algorithmic experimentation, modeling, system co-design, and benchmarking.

%% file: dlrm.bbl
\begin{thebibliography}{10}

\bibitem{bishop1995}
Christopher~M. Bishop.
\newblock {\em Neural Networks for Pattern Recognition}.
\newblock The Oxford University Press, 1st edition, 1995.

\bibitem{bonawitz2019federated}
Keith Bonawitz, Hubert Eichner, Wolfgang Grieskamp, Dzmitry Huba, Alex
  Ingerman, Vladimir Ivanov, Chloé Kiddon, Jakub Konečný, Stefano Mazzocchi,
  Brendan McMahan, Timon~Van Overveldt, David Petrou, Daniel Ramage, and Jason
  Roselander.
\newblock Towards federated learning at scale: System design.
\newblock In {\em Proc. 2nd Conference on Systems and Machine Learning
  (SysML)}, 2019.

\bibitem{cheng2016wide}
Heng-Tze Cheng, Levent Koc, Jeremiah Harmsen, Tal Shaked, Tushar Chandra,
  Hrishi Aradhye, Glen Anderson, Greg Corrado, Wei Chai, Mustafa Ispir, Rohan
  Anil, Zakaria Haque, Lichan Hong, Vihan Jain, Xiaobing Liu, and Hemal Shah.
\newblock Wide \& deep learning for recommender systems.
\newblock In {\em Proc. 1st Workshop on Deep Learning for Recommender Systems},
  pages 7--10, 2016.

\bibitem{Corinna1995}
Corinna Cortes and Vladimir~N. Vapnik.
\newblock Support-vector networks.
\newblock {\em Machine Learning}, 2:273--297, 1995.

\bibitem{Devroye1996}
Luc Devroye, Laszlo Gyorfi, and Gabor Lugosi.
\newblock {\em A Probabilistic Theory of Pattern Recognition}.
\newblock New York, Springer-Verlag, 1996.

\bibitem{duchi2011}
John Duchi, Elad Hazan, and Yoram Singer.
\newblock Adaptive subgradient methods for online learning and stochastic
  optimization.
\newblock {\em Journal of Machine Learning Research}, 12:2121--2159, 2011.

\bibitem{gloo}
Facebook.
\newblock Collective communications library with various primitives for
  multi-machine training (gloo),
  \url{https://github.com/facebookincubator/gloo}.

\bibitem{caffe2}
Facebook.
\newblock Caffe2, \url{https://caffe2.ai}, 2016.

\bibitem{frolov2017tensor}
Evgeny Frolov and Ivan Oseledets.
\newblock Tensor methods and recommender systems.
\newblock {\em Wiley Interdisciplinary Reviews: Data Mining and Knowledge
  Discovery}, 7(3):e1201, 2017.

\bibitem{gentry2009}
Craig Gentry.
\newblock A fully homomorphic encryption scheme.
\newblock PhD thesis, Stanford University, 2009.

\bibitem{golub1996}
Gene~H. Golub and Charles~F. Van~Loan.
\newblock {\em Matrix Computations}.
\newblock The John Hopkins University Press, 3rd edition, 1996.

\bibitem{commsref}
Ananth Grama, Vipin Kumar, Anshul Gupta, and George Karypis.
\newblock {\em Introduction to parallel computing}.
\newblock Pearson Education, 2003.

\bibitem{guo2017deepfm}
Huifeng Guo, Ruiming Tang, Yunming Ye, Zhenguo Li, and Xiuqiang He.
\newblock {DeepFM}: a factorization-machine based neural network for {CTR}
  prediction.
\newblock {\em arXiv preprint arXiv:1703.04247}, 2017.

\bibitem{hassan2007synthetic}
Rahman Hassan, Antony Harris, Nigel Topham, and Aris Efthymiou.
\newblock Synthetic trace-driven simulation of cache memory.
\newblock In {\em Proc. 21st International Conference on Advanced Information
  Networking and Applications Workshops (AINAW'07)}, 2007.

\bibitem{he2017neural}
Xiangnan He, Lizi Liao, Hanwang Zhang, Liqiang Nie, Xia Hu, and Tat-Seng Chua.
\newblock Neural collaborative filtering.
\newblock In {\em Proc. 26th Int. Conf. World Wide Web}, pages 173--182, 2017.

\bibitem{nccl}
Sylvain Jeaugey.
\newblock Nccl 2.0, 2017.

\bibitem{koren2009matrix}
Yehuda Koren, Robert Bell, and Chris Volinsky.
\newblock Matrix factorization techniques for recommender systems.
\newblock {\em Computer}, (8):30--37, 2009.

\bibitem{lian2018xdeepfm}
Jianxun Lian, Xiaohuan Zhou, Fuzheng Zhang, Zhongxia Chen, Xing Xie, and
  Guangzhong Sun.
\newblock {xDeepFM}: Combining explicit and implicit feature interactions for
  recommender systems.
\newblock In {\em Proc. of the 24th ACM SIGKDD International Conference on
  Knowledge Discovery \& Data Mining}, pages 1754--1763. ACM, 2018.

\bibitem{mlperf}
{MLPerf}.
\newblock \url{https://mlperf.org/}.

\bibitem{naumov2019}
Maxim Naumov.
\newblock On the dimensionality of embeddings for sparse features and data.
\newblock In {\em arXiv preprint arXiv:1901.02103}, 2019.

\bibitem{ning2015}
Xia Ning, Christian Desrosiers, and George Karypis.
\newblock A comprehensive survey of neighborhood-based recommendation methods.
\newblock In {\em Recommender Systems Handbook}, 2015.

\bibitem{mgp}
Pandora.
\newblock Music genome project \url{https://www.pandora.com/about/mgp}.

\bibitem{paszke2017pytorch}
Adam Paszke, Sam Gross, Soumith Chintala, and Gregory Chanan.
\newblock {PyTorch}: Tensors and dynamic neural networks in python with strong
  {GPU} acceleration \url{https://pytorch.org/}, 2017.

\bibitem{rendle2010}
Steffen Rendle.
\newblock Factorization machines.
\newblock In {\em Proc. 2010 IEEE International Conference on Data Mining},
  pages 995--1000, 2010.

\bibitem{sedhain2015autorec}
Suvash Sedhain, Aditya~Krishna Menon, Scott Sanner, and Lexing Xie.
\newblock {Autorec: Autoencoders meet collaborative filtering}.
\newblock In {\em Proc. 24th Int. Conf. World Wide Web}, pages 111--112, 2015.

\bibitem{walker1967}
Strother~H. Walker and David~B. Duncan.
\newblock Estimation of the probability of an event as a function of several
  independent variables.
\newblock {\em Biometrika}, 54:167--178, 1967.

\bibitem{wang2017deep}
Ruoxi Wang, Bin Fu, Gang Fu, and Mingliang Wang.
\newblock Deep \& cross network for ad click predictions.
\newblock In {\em Proc. ADKDD}, page~12, 2017.

\bibitem{zhou2018deepi}
Guorui Zhou, Na~Mou, Ying Fan, Qi~Pi, Weijie Bian, Chang Zhou, Xiaoqiang Zhu,
  and Kun Gai.
\newblock Deep interest evolution network for click-through rate prediction.
\newblock {\em arXiv preprint arXiv:1809.03672}, 2018.

\bibitem{zhou2018deep}
Guorui Zhou, Xiaoqiang Zhu, Chenru Song, Ying Fan, Han Zhu, Xiao Ma, Yanghui
  Yan, Junqi Jin, Han Li, and Kun Gai.
\newblock Deep interest network for click-through rate prediction.
\newblock In {\em Proc. of the 24th ACM SIGKDD International Conference on
  Knowledge Discovery \& Data Mining}, pages 1059--1068. ACM, 2018.

\end{thebibliography}
